\documentstyle[aps]{revtex}
\begin{document}
\draft
\title{Hierarchy Problem in the Shell-Universe Model} 
\author{Merab Gogberashvili}
\address{Institute of Physics, Tamarashvili st. 6, 380077 Tbilisi, Georgia \\ 
E-mail: gogber@hotmail.com} 
\date{\today}
\maketitle
\begin{abstract}
In the model where the Universe is considered as a thin shell expanding in
5-dimensional hyper-space there is a possibility to obtain  one scale for
particle theory corresponding to the 5-dimensional cosmological constant and
Universe thickness.
 \end{abstract}
\vskip 0.3cm

Several authors in the physics literature speculated about the possibility that our
Universe may be a thin membrane in a large dimensional hyper-Universe \cite{RS,V,S,BK}
(for simplicity here we consider the case of five dimensions). This approach is an
alternative to the conventional Kaluza-Klein picture that extra dimensions are
curled up to an unobservable size. In this paper we want to conceder Universe as a
bubble expanding in five dimensional space-time. The model of shell-Universe do not
contradict to present time experiments \cite{OW} and are supported by at least two
observed facts. First is the isotropic runaway of galaxies, which for close universe
model is usually explained as an expansion of a bubble in five dimensions. Second is
existence of preferred frame in the Universe where the relict background radiation
is isotropic. In the framework of the close-Universe model without boundaries this
can also be explained if the universe is 3-dimensional sphere and the mean velocity
of the background radiation is zero with respect to its center in the fifth
dimension. In shell-Universe models the expansion rate of the Universe should
depend not only on the matter density on the shell, but also on the properties of
matter in inner-outer regions. This can give rise to the effect similar to
the hypothetical dark matter. Also some authors want to introduce action
at a distance without ultrafast communication as a possible connection of
matter through the fifth dimension \cite{KH}.

In the case of non-compact extra dimensions one needs a mechanism to confine matter
inside the 4-dimensional manifold. Usually it is considered that trapping is
the result of the existence of special solution of 5-dimensional Einstein equations.
This trapping has to be gravitationally repulsive in nature and can be produced by a
large cosmological constant \cite{RS,V} or by some matter with negative energy
possibly filling the inner-outer space to the shell. In these models the zero-zero
component (the only component we need here) of the metric tensor slightly
generalize the standard Kaluza-Klein one 
\begin{equation}
\tilde{g}_{00} = \sigma (y) g_{00}(x^\mu)~~,
\label{1}
\end{equation}
where $x^\mu$ are ordinary coordinates of 4-dimensional space-time and $y$ is
the fifth coordinate. Trapping  is ruled by the potential
\begin{equation}
\sigma (y) \sim exp(E^2 |y|)~~,
\label{2}
\end{equation}
which is solution of Einstein equations and growth rapidly towards the fifth
dimension $y$. The integration constant $E^2$ (corresponding to the width of the
4-dimensional world $\epsilon$) is usually taken proportional to the 5-dimensional
cosmological constant
\begin{equation}
E^2 \sim \Lambda \sim \epsilon^{-2} ~~,
\label{3}
\end{equation}
to solve the cosmological constant problem in four dimensions. 

Indeed let us consider the Newton's approximation of 5-dimensional Einstein
equations with cosmological term for the trapped point-like source
\begin{equation}
(\Delta - \Lambda )\tilde{g}_{00} = 6\pi^2 G M \delta (r) \delta (y) ~~,
\label{4}
\end{equation}
where $\Delta$ is the 4-dimensional Laplacian. Using (\ref{2}) and (\ref{3}) and
separating variables we obtain the ordinary Newton's formula without the
cosmological term
\begin{equation}
g_{00} = 1 - 2gM/r ~~,
\label{5}
\end{equation}
where
\begin{equation}
g \sim G/\epsilon
\label{6}
\end{equation}
is the 4-dimensional gravitational constant.
   
It seems that the only new idea of the last decade in the subject of 
multidimensional models is an attempt to solve the hierarchy problem \cite{ADD}.
Hierarchy problem is a puzzle concerning masses of scalar fields. The light Higgs
with the mass $m_H \sim 10^3 GeV$ is needed in the Standard Model on the
electroweak scale. But masses of scalar fields quadratically divergent in
the loop expansion in quantum field theory and renormalization effects
should drive these masses up to the very large Plank scale 
\begin{equation}
M_P \sim g^{-1/2} \sim 10^{19} GeV ~~, 
\label{7}
\end{equation}
the natural cut-off of any quantum field theory.

In the case of higher-dimensional theories there is a possibility to obtain
very large Planck scale of the 4-dimensional world if the higher-dimensional
fundamental Planck scale is, for example, the gauge unification scale $m$.
Recently it was shown \cite{ADD} that in higher-dimensional models with
compact internal dimensions of size $\rho$, using the Newton's law in D+4
dimensions, at distances much larger than the size of internal dimensions  
\begin{equation}
M^2_P \sim \rho^D m^{2+D}~~ .
\label{8}
\end{equation}
It was claimed that this consideration can be physical in more than five
dimensions, since for $D = 1$ and $m \sim 10^3 GeV$ the size of internal
dimension will be huge $\rho \sim 10^{13} cm$.

We would like to notice here that this mechanism also takes place in models
with extended
extra dimensions. The only thing we need is to change the internal dimension radius
$\rho$ in (\ref{8}) by the Universe thickness $\epsilon$. Then the formula (\ref{6})
is just the same as (\ref{8}) for the case of five dimensions. Of course this
does not mean that the thickness of our world in the fifth dimension $\epsilon$
is the order of $\sim 10^{13} cm$. We must take into account that the measured
mass of the Higgs particle in four dimensions $m_H$ itself may be very 
different from the mass in five dimensions \cite{RS,V,S,BK} 
\begin{equation}
m^2_H \sim m^2 - \Lambda ~~ .
\label{9}
\end{equation}
For the case of $m^2 \sim \Lambda$ we can obtain any arbitrarily small value for
$m_H$, in particular $m_H \sim 10^3 GeV$ needed in the Standard Model. So even
in the 5-dimensional model we can have only one scale
\begin{equation}
m^3 \sim \Lambda^{3/2} \sim M^2_P/\Lambda \sim \epsilon^{-3}~~, 
\label{10}
\end{equation}
which corresponds to the thickness of the Universe.

The only parameter of the model $\Lambda$ can be measured, for example, in the
planned sub-millimeter measurements of gravity, since at distances of the
shell thickness size Newton's law must change. For these distances in the
right hand side of equation (\ref{4}) we have 4-dimensional delta function
$\delta(R)$, where $R$ is the 4-dimensional radial coordinate. So we can not
separate the variables and also hide the cosmological constant $\Lambda$. The
solution is
\begin{equation}
\tilde{g}_{00} = \frac{2}{z} \left[I_1(z) - \Lambda GM~K_1(z)\right]~~,
\label{11}
\end{equation}  
where $z= \Lambda^{1/2}R $  and $I_1, K_1$ are modified Bessel functions of order
one. In (\ref{11}) integration constants are chosen to give the Newton's
5-dimensional law 
\begin{equation}
\tilde{g}_{00} = 1 - 2GM/R^2
\label{12}
\end{equation}
in the limit $z \ll 1$.

So in the shell-universe model in five dimensions, except for solving some
principal problems of cosmology, there is a very attractive possibility to have
just the one fundamental scale corresponding to the thickness of the shell and
ruled by the 5-dimensional cosmological constant.  

\end{document}